\documentstyle[epsf]{prhep97}

\makeatletter
\let\chapter\hid@chapter
\makeatother

\def\alt{\mathrel{\mathop
  {\hbox{\lower0.5ex\hbox{$\sim$}\kern-0.8em\lower-0.7ex\hbox{$<$}}}}}
\def\agt{\mathrel{\mathop
  {\hbox{\lower0.5ex\hbox{$\sim$}\kern-0.8em\lower-0.7ex\hbox{$>$}}}}}

\begin{document}

\authorrunning{G.G.\,Raffelt}
\titlerunning{{\talknumber}: Astro-Particle Physics}
 
\def\talknumber{PL8} 

\title{\vbox to 0pt{\vskip-2.3cm\normalsize\tt\noindent 
To be published in the Proceedings of the
International Europhysics Conference on High Energy Physics 
(HEP 97),\protect\newline 
19-26 August 1997, Jerusalem, Israel \vfil}
\vskip-0.5cm\noindent
{\talknumber}: Astro-Particle Physics}
\author{Georg~G.\,Raffelt
(raffelt@mppmu.mpg.de)}
\institute{Max-Planck-Institut f\"ur Physik 
(Werner-Heisenberg-Institut)\\
F\"ohringer Ring 6, 80805 M\"unchen, Germany}

\maketitle

\begin{abstract}
Recent developments of those areas of astro-particle physics are
discussed that were represented at the HEP97 conference.  In
particular, the current status of direct and indirect dark-matter
searches and of TeV neutrino and $\gamma$-ray astronomy will be
reviewed.
\end{abstract}


\section{Introduction}

Astro-particle physics is such a wide field that it is certainly
impossible to review its current status in a single lecture. To make a
sensible selection it seemed most appropriate to cover those areas
which were represented in the parallel sessions of this conference,
i.e.\ mostly experimental topics in the areas of dark-matter detection
and of neutrino and $\gamma$-ray astronomy.  
One of the most cherished dark-matter candidates is the lightest
supersymmetric particle so that accelerator searches for supersymmetry
are of immediate cosmological importance, yet I consider this topic to
lie outside of my assignment. Likewise the laboratory searches for
neutrino masses and oscillations are of direct astrophysical and
cosmological significance, yet they exceed the boundaries of my
task. Finally, I will not cover the very exciting recent developments
in MeV to GeV neutrino astronomy (solar, supernova, and atmospheric
neutrinos) because they are reviewed by another speaker~\cite{PL11}.
 

\section{Dark Matter Searches}

\subsection{Dark Stars (MACHOs)}

The existence of huge amounts of dark matter in the universe is now
established beyond any reasonable doubt, but its physical nature
remains an unresolved mystery~\cite{Trimble,KolbTurner}.  A number of
well-known arguments negate the possibility of a purely baryonic
universe, but also point to significant amounts of nonluminous
baryons.  If some of them are in the galactic halo one most naturally
expects them to be in the form of Massive Astrophysical Compact Halo
Objects (MACHOs)---small and thus dim stars (brown dwarfs, M-dwarfs)
or stellar remnants (white dwarfs, neutron stars, black holes).
Stellar remnants and M-dwarfs are virtually excluded~\cite{Carr},
which leaves us with brown dwarfs, i.e.\ normal stars with a mass
below $0.08\,M_\odot$ (solar masses) so that they are too small to
ignite hydrogen.

Paczy\'nski proposed in 1986 to search for dim stars by the
``microlensing'' technique~\cite{Paczynski}.  A distant star brightens
with a characteristic lightcurve if a gravitational deflector passes
near the line of sight. Gravitational lensing produces two images, but
if their angular separation is too small the only observable effect is
the apparent brightening of the source.  A convenient sample of target
stars is provided by the Large Magellanic Cloud (LMC), a satellite
galaxy of the Milky Way. The LMC has enough bright stars and it is far
enough away and far enough above the galactic plane that one
intersects a good fraction of the galactic halo.  If MACHOs comprise
the halo, the lensing probability (``optical depth'' for microlensing)
is about $10^{-6}$ so that one has to monitor $\sim 10^{6}$ stars in
the LMC.  The duration of the brightness excursion depends on the lens
mass; for $1\,M_\odot$ it is typically 3~months, for
$10^{-2}\,M_\odot$ it is 9~days, for $10^{-4}\,M_\odot$ it is 1~day,
and for $10^{-6}\,M_\odot$ it is 2~hours.

\begin{figure}[b]
\centering\leavevmode
\epsfxsize=7.5cm
\epsfbox{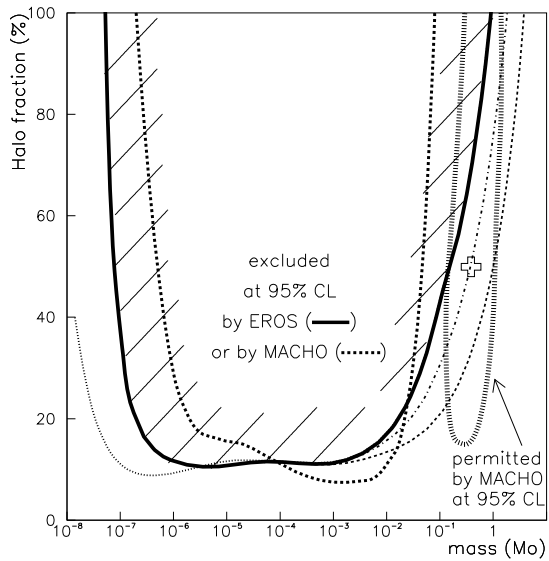}
\caption{Exclusion diagram at 95\% C.L.\ for the halo fraction and
mass of the assumed MACHOs~\protect\cite{EROS97}.  Their masses were
taken to be fixed and a standard model for the galactic halo was
used. The dotted line on the left is the EROS limit when blending and
finite size effects are ignored. The dot-dashed and dotted lines on
the right are the EROS limits when 1 or 2 of their events are
attributed to MACHOs. The cross is centered on the 95\% C.L.\
permitted range of the MACHO Collaboration~\protect\cite{Macho97}.
\label{fig:eros}}
\end{figure}

The microlensing search was taken up by the MACHO and the EROS
Collaborations, both reporting candidates toward the LMC since 1993
\cite{firstmachos}.  Moreover, the galactic bulge has been used as
another target where many more events occur through microlensing by
ordinary disk stars. While these observations are not sensitive to
halo dark matter, they allow one to develop a good understanding of
the microlensing technique and are an interesting method to study the
galaxy and its stellar content.  Within the past few years
microlensing has established itself as a completely new approach to
astronomy, with at least half a dozen collaborations pursuing
observations of various target regions. They also produce a huge
database of intrinsically variable stars, which is an invaluable
project in its own right.

Far from clarifying the status of dim stars as a galactic dark matter
contribution, the microlensing results toward the LMC with about a
dozen events~\cite{EROS97,Macho97} are quite confusing. For a standard
spherical halo the absence of short-duration events excludes a large
range of MACHO masses as a dominant component
(Fig.~\ref{fig:eros}).  On the other hand, the observed events
indicate a halo fraction between about 10\% and 100\% of MACHOs with
masses around $0.4\,M_\odot$ (Fig.~\ref{fig:eros}). This is
characteristic of white dwarfs, but a galactic halo consisting
primarily of white dwarfs is highly implausible and almost
excluded~\cite{Carr}.  Attributing the events to brown dwarfs ($M\alt
0.08\,M_\odot$) requires a very nonstandard density and/or velocity
distribution. Other explanations include an unexpectedly large
contribution from LMC stars, a thick galactic disk, an unrecognized
population of normal stars between us and the LMC, and other
speculations~\cite{LMCexplanations}. It is quite unclear which sort of
objects the microlensing experiments are seeing and where the lenses
are.

Meanwhile a first candidate has appeared in both the MACHO and EROS
data toward the Small Magellanic Cloud (SMC) \cite{SMC} which is
slightly more distant than the LMC and about $20^\circ$ away in the
sky. One event does not carry much statistical significance, but its
appearance is consistent with the LMC data if they are interpreted as
evidence for halo dark matter.  However, this interpretation would
imply a few solar masses for the SMC lens due to the large duration.

Besides more data from the LMC and SMC directions, other lines of
sight would be invaluable.  Of particular significance is the
Andromeda galaxy as a target because the line of sight cuts through
the halo almost vertically relative to the galactic disk.
Unfortunately, Andromeda is so far away that one cannot resolve
individual target stars.  One depends on the ``pixel lensing''
technique where one measures the brightening of a single pixel of the
CCD camera; one pixel covers the unresolved images of many stars. At
least two groups pursue this approach which has produced first
limits~\cite{pixel}.

\subsection{Axions}

While the microlensing searches seem to indicate that some
fraction of the galactic halo may consist of MACHOs, perhaps even in
the form of primordial black holes~\cite{PBH}, particle dark matter
{\em aficionados} should not get disheartened---weakly interacting
particles are still the best motivated option for the cold dark
matter which apparently dominates the universe.  

\begin{figure}[b]
\centering\leavevmode
\epsfxsize=9.2cm
\epsfbox{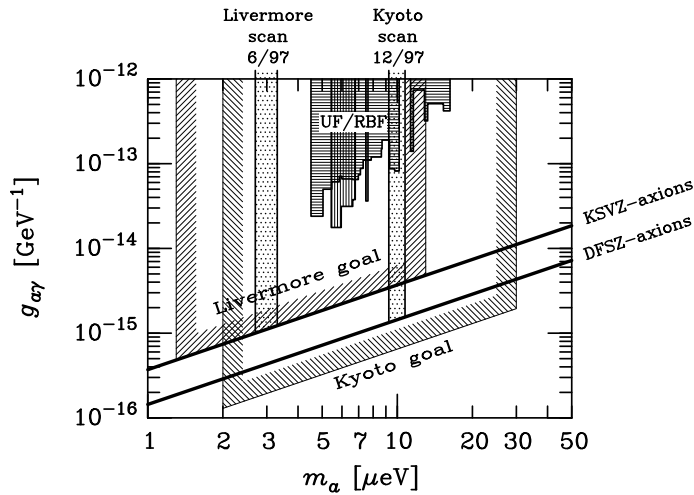}
\caption{Current limits on galactic dark matter axions from the
University of Florida (UF) \protect\cite{UFexperiment} and the
Rochester-Brookhaven-Fermilab (RBF) \protect\cite{RBFexperiment}
experiments and search goals of the Livermore~\protect\cite{Livermore}
and Kyoto~\protect\cite{Kyoto} experiments. It was assumed that the
local galactic axion density is $300\,\rm MeV\,cm^{-3}$. The
axion-photon coupling is given by ${\cal L}_{\rm int}=g_{a\gamma} {\bf
E}\cdot{\bf B}\,a$. The relationship between $g_{a\gamma}$ and $m_a$
for the popular DFSZ and KSVZ models is indicated.
\label{fig:axion}}
\end{figure}

One of two well motivated possibilities are axions which appear as
Nambu-Goldstone bosons of the spontaneously broken Peccei-Quinn
symmetry which is motivated as a solution of the CP problem of strong
interactions~\cite{PecceiQuinn}. Apart from numerical parameters of
order unity, these models are characterized by a single unknown
quantity, the Peccei-Quinn scale $f_a$ or the axion mass
$m_a=0.62\,{\rm eV}\,(10^7\,{\rm GeV}/f_a)$. In the early universe
axions form nonthermally as highly occupied and thus quasi-classical
low-momentum oscillations of the axion field.  If axions are the dark
matter, a broad class of early-universe scenarios predicts $m_a$ to
lie in the range $1\,\mu{\rm eV}$ to $1\,{\rm meV}$ \cite{Davis}.

In a magnetic field axions convert into photons by the Primakoff
process because they have a two-photon coupling~\cite{Sikivie}.  A
frequency of $1\,\rm GHz$ corresponds to $4\,\rm \mu eV$; a search
experiment for galactic axions consists of a high-$Q$ microwave
resonator placed in a strong magnetic field.  At low temperature one
looks for the appearance of microwave power beyond thermal and
amplifier noise.  Two pilot
experiments~\cite{UFexperiment,RBFexperiment} could not reach
realistic axion models, but two ongoing experiments with much larger
cavity volumes have the requisite sensitivity (Fig.~\ref{fig:axion}).
In its current setup, the Livermore experiment~\cite{Livermore} uses
conventional microwave amplifiers while the Kyoto
experiment~\cite{Kyoto} employs a completely novel detection technique
based on the excitation of a beam of Rydberg atoms which passes
through the cavity.

\subsection{Weakly Interacting Massive Particles (WIMPs)}

The other favored class of particle dark matter candidates are WIMPs,
notably the lightest supersymmetric particles in the form of
neutralinos~\cite{JKG96}. Direct searches rely on WIMP-nucleus
scattering, for example in Ge or NaI crystals~\cite{Goodman}.  The
expected counting rate is of order $1~\rm event~kg^{-1}~day^{-1}$ and
thus extremely small.  To beat cosmic-ray and radioactive backgrounds
one must go deeply underground and use ultra pure materials. The
recoils for 10--$100\,\rm GeV$ WIMP masses are of order $10\,\rm
keV$. Such small energy depositions can be measured by electronic,
bolometric, and scintillation techniques. The number of experimental
projects is too large to even list them here~\cite{DMReviews}.

\begin{figure}[b]
\vskip-0.5cm
\centering\leavevmode
\epsfxsize=8.0cm
\epsfbox{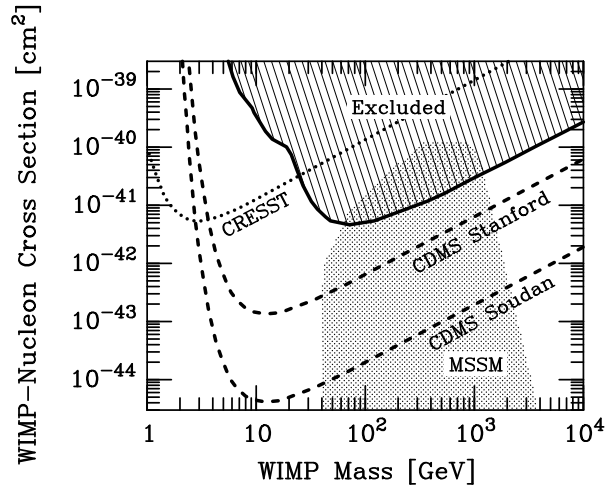}
\caption{Exclusion range for the spin-independent WIMP scattering
cross section per nucleon from the NaI
experiments~\protect\cite{UKDMC,DAMA96} and the germanium
detectors~\protect\cite{Germanium}. Also shown is the range of
expected counting rates for dark-matter neutralinos in the minimal
supersymmetric standard model (MSSM) without universal scalar mass
unification~\protect\cite{Gondolo,Bottino}. The search goals for the
upcoming large-scale cryogenic experiments
CRESST~\protect\cite{CRESST} and CDMS~\protect\cite{CDMS} are also
shown, where CDMS is located at a shallow site at Stanford, but will
improve its sensitivity after the planned move to a deep site in the
Soudan mine.
\label{fig:wimps}}
\end{figure}

The current limits already dig into the supersymmetric parameter space
(Fig.~\ref{fig:wimps}).  The DAMA/NaI experiment has actually reported
a WIMP signature~\cite{DAMA97} which would point to neutralinos just
below their previous exclusion range~\cite{Bottino97}.  The
significance of this result is very low, and tentative signals are
bound to appear just below the previous exclusion range. Still, the
good news is that this detection {\em could\/} be true in the sense
that one has reached the sensitivity necessary to find supersymmetric
dark matter.  In the near future the large-scale cryogenic detectors
CRESST~\cite{CRESST} and CDMS~\cite{CDMS} will explore a vast space of
WIMP-nucleon cross-sections (Fig.~\ref{fig:wimps}).


\section{Neutrino Astronomy}

Indirect methods to search for WIMPs rely on their annihilation in the
galactic halo or in the center of the Sun or Earth where WIMPs can be
trapped~\cite{JKG96}.  The search for GeV--TeV neutrinos from the Sun
or Earth in the Kamiokande, Baksan, and MACRO
detectors~\cite{tele-obs} already touch the parameter range relevant
for neutralino dark matter~\cite{tele-theory}.  Neutrino telescopes
are thus competitive with direct dark-matter searches, where it
depends on details of the supersymmetric models which approach has a
better chance of finding neutralinos.  Roughly, a water or ice
Cherenkov detector requires a $\rm km^3$ volume to be competitive with
the CDMS-Soudan search goal.

After the sad demise of the deep-sea DUMAND project a $\rm km^3$
neutrino telescope has again come within realistic reach after the
breathtaking progress in the development of the AMANDA ice Cherenkov
detector at the south pole~\cite{AMANDA}. The lake Baikal water
Cherenkov detector~\cite{Baikal} is another operational neutrino
telescope, but probably it cannot reach the $\rm km^3$ size. The
prospects of the deep-sea projects NESTOR~\cite{NESTOR} and
ANTARES~\cite{ANTARES} in the Mediterranean depend on the outcome of
their current ``demonstrator'' phase. Either way, after the explosive
development of solar and atmospheric neutrino observatories (MeV--GeV
energies), high-energy neutrino astronomy is set to become a reality
in the very near future.

Besides the search for dark matter, neutrino astronomy addresses
another old and enigmatic astrophysical problem, the origin of cosmic
rays which engulf the Earth with energies up to $\sim 10^{20}\,\rm
eV$.  They consist of protons and nuclei which must be accelerated
somewhere in the universe. Whenever they run into stuff they produce
pions and thus neutrinos and photons in roughly equal proportions
(``cosmic beam dumps''). Because the universe is opaque to photons
with energies exceeding a few $10\,\rm TeV$ due to pair production on
the cosmic microwave background, and because charged particles are
deflected by magnetic fields, high-energy neutrino astronomy offers a
unique observational window to the universe, and especially a chance
to identify the sites of cosmic-ray acceleration~\cite{Gaisser}.


\section{TeV $\gamma$-Ray Astronomy}

Perhaps the most attractive sites for the cosmic-ray acceleration are
active galactic nuclei (AGN) which are likely powered by accreting
black holes.  These objects tend to eject huge jets in opposite
directions; for an estimate of the expected neutrino flux see
Ref.~\cite{Zas}. If this is indeed the case one would equally expect
high-energy $\gamma$-rays from these sources. Remarkably, for the past
few years TeV $\gamma$-rays have indeed been
observed~\cite{mrk421,mrk501,Hegra97} from the two nearby ($\sim 300$
million light-years) AGNs Markarian 421 and 501 which have jets
pointing toward Earth.

\begin{figure}[b]
\centering\leavevmode
\epsfxsize=8.1cm
\epsfbox{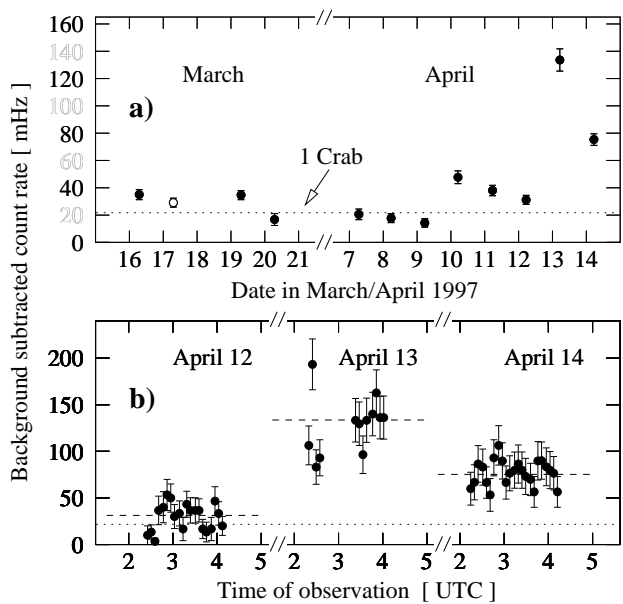}
\caption{Detection rate of the active galaxy Mrk~501 in the HEGRA
stereoscopic system of Imaging Air Cherenkov Telescopes on a night by
night basis, {\bf a)}~for the whole data set and {\bf b)}~for the last
3 nights in 5 min.\ intervals~\protect\cite{Hegra97}.  The dashed
lines indicate the average per night, the dotted line shows the Crab
detection rate as a reference.  Errors are statistical only.
\label{fig:hegra}}
\end{figure}

Until recently $\gamma$-ray astronomy reached only up to $\sim 20\,\rm
GeV$ because the low fluxes at higher energies require forbiddingly
large satellites. The observational break-through in the TeV range
arose from Imaging Air Cherenkov Telescopes (IACTs) on the
ground~\cite{Jelley}. A high-energy $\gamma$-ray hits the upper
atmosphere at an altitude of $\sim 16\,\rm km$ and produces an
electromagnetic shower which in turn produces Cherenkov light.  With a
relatively crude telescope one can thus take an image of the
shower. The axes of the cigar-shaped shower images of many
$\gamma$-rays intersect in one point which corresponds to the location
of the source in the sky and thus allows one to discriminate against
the much larger but isotropic flux of hadronic cosmic rays. A number
of galactic sources are now routinely observed, notably the Crab
nebula, which is seen at energies up to $50\,\rm TeV$ and serves as a
``calibration'' source.  There remains an unexplored spectral range
between about 20 and $300\,\rm GeV$ which requires much larger IACTs
than are currently available.

The Markarians are the first extragalactic sources in the TeV
$\gamma$-sky. Their behavior is quite tantalizing in that they are
hugely variable on sub-hour time scales
(Fig.~\ref{fig:hegra}). Moreover, Mrk~501 essentially ``switched on''
from a state of low activity with about 0.1 of the Crab flux in 1995,
about 0.3 in 1996, to about 10 times the Crab since February 1997.  If
the high-energy photons are produced by a proton beam or, say, photon
upscattering by accelerated electrons is by no means obvious.  One
will need neutrino telescopes to decide this question and thus to
understand these intruiging objects.


\section{Summary}

Experiments to identify the physical nature of the galactic dark
matter have recently made great strides.  The microlensing searches
for MACHOs have observed around a dozen candidates toward the Large
Magellanic Cloud, but an interpretation of these events is quite
puzzling in that their best-fit mass puts them into the white-dwarf
category which is highly implausible. This year a first candidate
toward the Small Magellanic Cloud has been reported, with an even
larger apparent mass.  From virtually any cosmological perspective
cold dark matter remains the favored hypothesis for the dominant mass
fraction of the universe.  Full-scale searches for the most favored
particle candidates (axions and WIMPs) are in progress.  One can hunt
WIMPs also by searching for their annihilation products in the form of
high-energy neutrinos from the Sun or the center of the Earth.  With a
$\rm km^3$ water or ice Cherenkov detector one could cover a
significant fraction of the parameter space for supersymmetric dark
matter. The impressive progress of the AMANDA south pole detector and
the appearance of new deep-sea projects (ANTARES and NESTOR) bode well
for this approach---a $\rm km^3$ detector could be up and running
before the LHC.  High-energy neutrino astronomy has other intruiging
objectives, notably searching for the sites of cosmic-ray
acceleration. The breathtaking recent observations of TeV
$\gamma$-rays from the two nearby active galaxies Markarian 421 and
501 have zoomed this physics target into sharper focus. Whether or not
active galactic nuclei accelerate protons is a question that can be
answered only by neutrino astronomy.


\subsection*{Acknowledgments}

This work was supported, in part, by the Deutsche
Forschungsgemeinschaft under grant No.\ SFB~375.



\begin{thebibliography}{99}

\bibitem{PL11}
  M.~Nakahata, PL11, these Proceedings.                         

\bibitem{Trimble}
  V.~Trimble, Annu. Rev. Astron. Astrophys. 25 (1987) 425.
  S.~Tremaine, Physics Today 45:2 (1992) 28.  

\bibitem{KolbTurner}
  E.W.~Kolb and M.S.~Turner, {\em The Early Universe}
  (Addison-Wesley, Redwood City, 1990). 
  G.~B\"orner, {\em The Early Universe}, 2nd edition
  (Springer, Berlin, 1992).

\bibitem{Carr} 
  B.J.~Carr, Comments Astrophys. 14 (1990) 257;
  Annu. Rev. Astron. Astrophys. 32 (1994) 531.
  B.J.~Carr and M.~Sakellariadou, 
  Fermilab-Pub-97-299-A, Astrophys. J. (submitted 1997).
  D.J.~Hegyi and K.A.~Olive, Astrophys. J. 303 (1986) 56. 

\bibitem{Paczynski}
  B.~Paczy\'nski, Astrophys. J. \textbf{304} (1986) 1.

\bibitem{firstmachos}
  C.~Alcock et al. (MACHO Collab.),
  Nature 365 (1993) 621;
  Phys. Rev. Lett. 74 (1995) 2867.
  E.~Aubourg et al. (EROS Collab.),
  Nature 365 (1993) 623;
   Astron. Astrophys. 301 (1995) 1.
  R.~Ansari et al. (EROS Collab.),
  Astron. Astrophys. 314 (1996) 94.

\bibitem{EROS97}
  C.~Renault et al. (EROS Collab.),
  Astron. Astrophys. 324 (1997) L69.

\bibitem{Macho97}
  C.~Alcock et al. (MACHO Collab.),
  Astrophys. J. 471 (1996) 774; 
  Astrophys. J. 486 (1997) 697.

\bibitem{LMCexplanations}
  K.~Sahu, Nature 370 (1994) 275.
  H.S.~Zhao, astro-ph/9606166, 9703097. 
  C.~Alcock et al. (MACHO Collab.), Astrophys. J. 490 (1997) L59.
  A. Gould, astro-ph/9709263.
  N. Evans et al., astro-ph/9711224.
  E. Gates et al., astro-ph/9711110.
  D.~Zaritsky and D.~Lin, astro-ph/9709055.

\bibitem{SMC}
  C.~Alcock et al. (MACHO Collab.), Astrophys. J. 491 (1997) L11.
  N.~Palanque-Delabrouille (EROS Collab.), astro-ph/9710194. 

\bibitem{pixel}
  A.P.S.~Crotts and A.B.~Tomaney, Astrophys. J. 473 (1996) L87.
  R.~An\-sari et al. (AGAPE Collab.), 
  Astron. Astrophys. 324 (1997) 843.

\bibitem{PBH}
  J.~Yokoyama, Astron. Astrophys. 318 (1997) 673.
  K.~Jedamzik, Phys. Rev. D 55 (1997) R5871.
  J.C.~Niemeyer and K.~Jedamzik,
  astro-ph/ 9709072.

\bibitem{PecceiQuinn}
  R.D.~Peccei and H.R.~Quinn, Phys. Rev. Lett. 38 (1977) 1440; 
  Phys. Rev. D 16 (1977) 1791. 
  S.~Weinberg, Phys. Rev. Lett. 40 (1978) 223.
  F.~Wilczek, Phys. Rev. Lett. 40 (1978) 279. 
  J.E.~Kim, Phys. Rept. 150 (1987) 1.
  H.-Y.~Cheng, Phys. Rept. 158 (1988) 1.

\bibitem{Davis}
  R.L.~Davis, Phys. Lett. B 180 (1986) 225. 
  R.A.~Battye and E.P.S. Shellard, 
  Phys. Rev. Lett. 73 (1994) 2954;
  (E) ibid. 76 (1996) 2203. 
  D.~Harari and P.~Sikivie, Phys. Lett. B 195 (1987) 361.
  C.~Hagmann and P.~Sikivie, Nucl. Phys. B 363 (1991) 247. 

\bibitem{Sikivie}
  P.~Sikivie, Phys. Rev. Lett. 51 (1983) 1415;
  Phys. Rev. D 32 (1985) 2988. 

\bibitem{UFexperiment}
  C.~Hagmann, P. Sikivie, N.S. Sullivan, D.B.~Tanner,  
  Phys. Rev. D 42 (1990) 1297.

\bibitem{RBFexperiment}
  W.U.~Wuensch et al., Phys. Rev. D 40 (1989) 3153.

\bibitem{Livermore}
  C.~Hagmann et al.,   Nucl. Phys. Proc. Suppl. 51B (1996) 209. 

\bibitem{Kyoto} 
  I.~Ogawa, S.~Matsuki and K.~Yamamoto,
  Phys. Rev. D 53 (1996) R1740. 

\bibitem{JKG96}
  G.~Jungman, M.~Kamionkowski and K.~Griest,
  Phys. Rept. 267 (1996) 195.

\bibitem{Goodman}
  M.W.~Goodman and E.~Witten, Phys. Rev. D 31 (1985) 3059. 

\bibitem{DMReviews}
  J.~Primack, D.~Seckel and B.~Sadoulet,
  Annu. Rev. Nucl. Part. Sci. 38 (1988) 751.
  P.F.~Smith and J.D.~Lewin, Phys. Rept. 187 (1990) 203.
  D.O.~Caldwell, Mod. Phys. Lett. A 5 (1990) 1543;
  Proc. TAUP97, to be published.
  N.E.~Booth, B.~Cabrera and E.~Fiorini,
  Annu. Rev. Nucl. Part. Sci. 46 (1996) 471.
  A.~Watson, Science 275 (1997) 1736.

\bibitem{UKDMC}
  P.F.~Smith et al., Phys. Lett. B 379 (1996) 299. 
  J.J.~Quenby, Astropart. Phys. 5 (1996) 249. 

\bibitem{DAMA96}
  R.~Bernabei et al., Phys. Lett. B 389 (1996) 757.

\bibitem{Germanium}
  D.~Reusser et al., Phys. Lett. B 255 (1991) 143. 
  E.~Garcia et al., Phys. Rev. D 51 (1995) 1458.  
  M.~Beck et al., Phys. Lett. B 336 (1994) 141.

\bibitem{Gondolo}
  P.~Gondolo, private communication (1997). 

\bibitem{Bottino}
  L.~Bergstr\"om and P.~Gondolo,
  Astropart. Phys. 5 (1996) 263. 
  A.~Bottino, F.~Donato, G.~Mignola and S.~Scopel,
  Phys. Lett. B 402 (1997) 113. 
  J.~Edsj\"o and P.~Gondolo,
  Phys. Rev. D 56 (1997) 1879.

\bibitem{CRESST} 
  M.~Sisti et al., in: Proc. 7th Int. Workshop on Low
  Temperature Detectors (LTD-7), 27 July--2 August 1997, Munich,
  Germany, ed. by S.~Cooper (Max-Planck-Institut f\"ur
  Physik, Munich, 1997).

\bibitem{CDMS}
  S.W.~Nam et al., in: Proc. 7th Int. Workshop on Low Temperature 
  Detectors (LTD-7), 27 July - 2 August 1997, Munich, Germany, ed. 
  by S.~Cooper (Max-Planck-Institut f\"ur 
  Physik, Munich, 1997).

\bibitem{DAMA97}
  R.~Bernabei et al., astro-ph/9710290,
  to be published in Proc. TAUP97. 

\bibitem{Bottino97}
  A.~Bottino, F.~Donato, N.~Fornengo and S.~Scopel,
  astro-ph/9709292 and 9710295. 

\bibitem{tele-obs} 
  M.~Mori et al. (Kamiokande Collab.), Phys. Rev. D 48 (1993) 5505.  
  M.M.~Boliev et al. (Baksan telescope), in: Proc. TAUP95,
  ed. by A.~Morales, Nucl. Phys. B (Proc. Suppl.) 48 (1996) 83.
  M.~Ambrosio et al. (MACRO Collab.), Preprint INFN-AE-97-23.
  
\bibitem{tele-theory}
  M.~Kamionkowski et al., Phys. Rev. Lett. 26 (1995) 5174.
  V.~Berezinskii, et al., 
  Astropart. Phys. 5 (1996) 333. 
  L.~Bergstr\"om, J.~Edsj\"o and P.~Gondolo, 
  Phys. Rev. D 55 (1997) 1765.

\bibitem{AMANDA}
  L.~Berstr\"om, astro-ph/9612122,
  Proc. Identification of Dark Matter,
  Sheffield, UK, Sept. 1997. 
  F.~Halzen, astro-ph/9707289. 

\bibitem{NESTOR}
  E.G.~Anassontzis et al. (NESTOR Collab.)
  Preprint DFF-283-7-1997, Jul 1997.
  Proc. 18th International Symposium on Lepton-Photon 
  Interactions (LP 97), Hamburg, Germany, 28 July--1 Aug. 1997. 

\bibitem{ANTARES}
  ANTARES Collaboration, Proposal, astro-ph/9707136.
  
\bibitem{Baikal}
  V.A.~Balkanov (Baikal Collab.), astro-ph/9705017,
  Proc. XXXII Rencontres de Moriond, Les Arcs,
  France, Jan. 18--25, 1997. 

\bibitem{Gaisser}
  T.K.~Gaisser, astro-ph/9707283,
  Proc. OECD Megascience Forum Workshop, Taormina, Italy,
  May 22--23, 1997. 
  F.~Halzen, astro-ph/9605014, 
  Proc. Venitian Neutrino Conference, Feb. 1996.

\bibitem{Zas}
  F.~Halzen and E.~Zas, Astrophys. J. 488 (1997) 669.

\bibitem{mrk421}
  M.~Punch et al., Nature 358 (1992) 477.
  A.D.~Kerrick et al., Astrophys. J. 438 (1995) L59.
  D.~Petry et al., Astron. Astrophys. 311 (1996) L13.  

\bibitem{mrk501}
  J.~Quinn et al., Astrophys. J. 456 (1996) L83.

\bibitem{Hegra97}
  F.~Aharonian et al. (HEGRA Collab.),
  Astron. Astrophys. 327 (1997) L5.
  M.~Catanese et al., astro-ph/9707179.


\bibitem{Jelley}
  J.V.~Jelley and T.C.~Weekes,
  Sky \& Telescope, Sept.~1995, pg.20. 

\end{thebibliography}
\end{document}